\documentclass{ws-procs9x6}

\begin{document}

\title{SPIN QUANTUM PLASMAS --- NEW ASPECTS OF COLLECTIVE DYNAMICS}
\author{M. MARKLUND$^*$  and G. BRODIN}

\address{Department of Physics, Ume\aa\ University, SE-901 87 Ume\aa, Sweden \\
$^*$E--mail: mattias.marklund@physics.umu.se}

\begin{abstract}
Quantum plasmas is a rapidly expanding field of research, with applications
ranging from nanoelectronics, nanoscale devices and ultracold plasmas,
to inertial confinement fusion and astrophysics. Here we give a short systematic
overview of quantum plasmas. In particular, we analyze the collective
effects due to spin using fluid models. The introduction
of an intrinsic magnetization due to the plasma electron (or positron) spin properties in the magnetohydrodynamic limit is discussed. Finally, a discussion of the theory and examples of 
applications is given. 
\end{abstract}

\bodymatter

\section{Introduction}
The field of quantum plasmas is a rapidly growing field of research.
From the non-relativistic domain, with its basic description in terms 
of the Schr\"odinger equation, to the strongly relativistic regime, with its
natural connection to quantum field theory, quantum plasma physics
provides promises of highly interesting and important application,
fundamental connections between different areas of science, as well
as difficult challenges from a computational perspective. The necessity to 
thoroughly understand such plasmas motivates a reductive principle of
research, for which we successively build more complex models based
on previous results. The simplest lower order effect due to 
relativistic quantum mechanics is the introduction of spin, and as such thus provides a first step towards a partial description of relativistic quantum plasmas. 

Already in the 1960's, Pines studied the excitation spectrum of quantum
plasmas \cite{Pines,pines-book}, for which we have a high density and a low temperature
as compared to normal plasmas. In such systems, the finite width of the
electron wave function makes quantum tunnelling effects crucial, leading to
an altered dispersion relation. Since the pioneering work by Pines, a
number or theoretical studies of quantum statistical properties of plasmas
has been done (see, \textit{e.g.}, Ref.\ \refcite{kremp-etal} and references therein). For
example, Bezzerides \& DuBois presented a kinetic theory for the quantum
electrodynamical properties of nonthermal plasmas \cite{bezzerides-dubois},
while Hakim \& Heyvaerts presented a covariant Wigner function approach for
relativistic quantum plasmas \cite{hakim-heyvaerts}. Recently there has been
an increased interest in the properties of quantum plasmas \cite
{Manfredi2005,haas-etal1,haas,shukla,garcia-etal,collection2C,collection2E,collection2G,collection2H,collection2I,collection2K,collection2L,Haas-HarrisSheet,marklund-brodin,brodin-marklund,BM-pairplasma,%
shukla-eliasson,shukla-eliasson2,shaikh-shukla,brodin-marklund2}%
. The studies has been motivated by the development in nanostructured
materials \cite{craighead} and quantum wells \cite{manfredi-hervieux}, the
discovery of ultracold plasmas \cite{li-etal} (see Ref.\ \refcite{fletcher-etal}
for an experimental demonstration of quantum plasma oscillations in Rydberg
systems), astrophysical applications \cite{harding-lai}, or a general
theoretical interest. Moreover, it has recently been experimentally shown
that quantum dispersive effects are important in inertial confinement
plasmas \cite{glenzer-etal}. The list of quantum mechanical effects that can
be included in a fluid picture includes the dispersive particle properties
accounted for by the Bohm potential \cite
{Manfredi2005,haas-etal1,haas,shukla,garcia-etal,collection2C,collection2E,collection2G,collection2H,collection2I,collection2K,collection2L,Haas-HarrisSheet}, the zero temperature
Fermi pressure \cite{Manfredi2005,haas-etal1,haas,shukla,garcia-etal}, spin
properties \cite{marklund-brodin,brodin-marklund,BM-pairplasma} as well as
certain quantum electrodynamical effects \cite
{marklund-shukla,Lundin2007,lundstrom-etal,Brodin-etal-2007}. Within such descriptions, \cite
{Manfredi2005,haas-etal1,haas,shukla,garcia-etal,marklund-brodin,brodin-marklund,Lundin2007,lundstrom-etal,Brodin-etal-2007}
 quantum and classical collective effects can be described within a unified
picture.

\section{The microscopic equations: Schr\"odinger and Pauli dynamics}

\subsection{The Schr\"odinger description}

The basic equation of nonrelativistic quantum mechanics is the Schr\"odinger
equation. The dynamics of an electron, represented by its wave function $\psi$, in an external electromagnetic potential
$\phi$ is governed by 
\begin{equation}\label{eq:schrodinger}
  i\hbar\frac{\partial \psi}{\partial t} + \frac{\hbar}{2 m_e}\nabla^2\psi + e\phi\psi  = 0 ,
\end{equation}
where $\hbar$ is Planck's constant, $m_e$ is the electron mass, and $e$ is the magnitude
of the electron charge. This complex equation may be written as two real equations, writing 
$\psi = \sqrt{n}\,\exp{iS/\hbar}$, where $n$ is the amplitude and $S$ the phase of the wave function, respectively \cite{holland}. Such a decomposition was presented by de Broglie and Bohm in order to 
understand the dynamics of the electron wave packet in terms of classical variables. 
Using this decomposition in Eq.\ (\ref{eq:schrodinger}), we obtain
\begin{equation}\label{eq:schrod-cont}
  \frac{\partial n}{\partial t} + \nabla\cdot(n\mathbf{v}) = 0 , 
\end{equation}
and 
\begin{equation}\label{eq:schrod-mom}
  m_e\frac{d\mathbf{v}}{d t} = e\nabla\phi
    + \frac{\hbar^2}{2m_e}\nabla\left(\frac{\nabla^2\sqrt{n}}{\sqrt{n}}\right) ,
\end{equation}
where the velocity is defined by $\mathbf{v} = \nabla S/m_e$. The last term of Eq.\ (\ref{eq:schrod-mom})
is the gradient of the Bohm--de Broglie potential, and is due to the effect of wave function spreading, giving rise to a dispersive-like term. We also note the
striking resemblance of Eqs.\ (\ref{eq:schrod-cont}) and (\ref{eq:schrod-mom}) to the classical fluid equations.

\subsection{The Pauli description}

In relativistic quantum mechanics, the spin of the electron (and positron) 
is rigorously introduced through the Dirac Hamiltonian 
\begin{equation}\label{eq:dirac}
  H = c\mathbf{\alpha}\cdot\left( \mathbf{p} + e\mathbf{A} \right) - e\phi + \beta m_ec^2 ,
\end{equation}
where $\mathbf{\alpha} = (\alpha_1, \alpha_2, \alpha_3)$, $e$ is the magnitude of the electron charge,
$c$ is the speed of light, $\mathbf{A}$ is the vector potential, $\phi$ is the electrostatic potential,
and the relevant matrices are given by
\begin{equation}
  \mathbf{\alpha} = \left( 
    \begin{array}{cc}
      0                 & \mathbf{\sigma} \\
      \mathbf{\sigma} & 0
    \end{array}
  \right) , \qquad
  {\beta} = \left( 
    \begin{array}{cc}
      {\mathsf{I}} & 0 \\
      0                     & -{\mathsf{I}}
    \end{array}
  \right) .
\end{equation}
Here ${\mathsf{I}}$ is the unit $2\times 2$ matrix and $\mathbf{\sigma} = (\sigma_1,\sigma_2, \sigma_3)$, where we have the Pauli spin matrices 
\begin{equation}
\sigma_1 = \left( 
\begin{array}{cc}
0 & 1 \\ 
1 & 0
\end{array}
\right) , \, \sigma_2 = \left( 
\begin{array}{cc}
0 & -i \\ 
i & 0
\end{array}
\right) , \, \text{ and }\, \sigma_3 = \left( 
\begin{array}{cc}
1 & 0 \\ 
0 & -1
\end{array}
\right) .
\end{equation}
From the Hamiltonian (\ref{eq:dirac}), a nonrelativistic counterpart may be obtained, 
taking the form
\begin{equation}\label{eq:pauli-hamiltonian}
  H = \frac{1}{2m_e}\left( \mathbf{p} + e\mathbf{A} \right)^2 
    + \frac{e\hbar}{2m_e}\mathbf{B}\cdot\mathbf{\sigma} - e\phi .
\end{equation}
Thus, the electron possesses a magnetic moment $\mathbf{m} = 
-\mu_B\langle\psi|\mathbf{\sigma}|\psi\rangle/\langle\psi|\psi\rangle$, where 
$\mu_B = e\hbar/2m_e$ is the Bohr magneton, giving a contribution $-\mathbf{B}\cdot\mathbf{m}$
to the energy. The latter shows the paramagnetic property of the electron, where the spin vector is
anti-parallel to the magnetic field in order to 
minimize the energy of the magnetized system.  
According to (\ref{eq:pauli-hamiltonian}) and the relation $dF/dt = \partial F/\partial t + (1/i\hbar)[F, H]$,
where $F$ is some operator and $[,]$ is the Poisson bracket, we have the following
evolution equations for the position and momentum in the Heisenberg picture
\cite{holland,Dirac,Barut-Thacker}
\begin{equation}
  \mathbf{v} \equiv \frac{d\mathbf{x}}{dt} = \frac{1}{m_e}\left( \mathbf{p} + e\mathbf{A} \right) ,
\end{equation}
\begin{equation}
  m_e\frac{d\mathbf{v}}{dt} = -e\left( \mathbf{E} + \mathbf{v}\times\mathbf{B} \right) 
    - \frac{2}{\hbar}\mu_B\mathbf{\nabla}(\mathbf{B}\cdot\mathbf{s}) ,
\end{equation}
while the spin evolution is given by 
\begin{equation}
  \frac{d\mathbf{s}}{dt} = \frac{2}{\hbar}\mu_B \mathbf{B}\times\mathbf{s}, 
\end{equation}
where the spin operator is given by 
\begin{equation}
  \mathbf{s} = \frac{\hbar}{2}\mathbf{\sigma} .
\end{equation}
The above equations thus gives the quantum operator equivalents of the 
equations of motion for a classical particle, including the evolution of the
spin in a magnetic field. 

The non-relativistic evolution of spin $\tfrac{1}{2}$ particles, as
described by the two-component spinor $\varPsi_{(\alpha)}$, is given by
the Pauli equation (see, \textit{e.g.}, \refcite{holland}) 
\begin{equation}  \label{eq:pauli}
i\hbar\frac{\partial\psi}{\partial t} + \left[ \frac
{\hbar^2}{2m_e}\left(\nabla + \frac{ie}{\hbar}%
\mathbf{A} \right)^2 - \mu_B\mathbf{B}\cdot\mathbf{\sigma} + e\phi %
\right] \psi = 0 ,
\end{equation}
where $\mathbf{A}$ is the vector potential, 
$\mu_B = e\hbar/2m_e$ is the Bohr magneton,  
and $\mathbf{\sigma} = (\sigma_1, \sigma_2, \sigma_3)$ is the Pauli spin
vector. 

Now, in the same way as in the Schr\"odinger case, we may decompose 
the electron wave function $\psi$ into its amplitude and phase. However, as
the electron has spin, the wave function is now represented by a 2-spinor
instead of a c-number. Thus, we may use $\psi = \sqrt{n}\,\exp(iS/\hbar)\varphi$,
where $\varphi$, normalized such that $\varphi^{\dag}\varphi = 1$, now gives the spin part of the wave function.
 Multiplying the Pauli equation (\ref{eq:pauli}) by $\psi^{\dag}$, inserting the above wave function 
 decomposition
and taking the gradient of the resulting phase evolution equation, we obtain
the conservation equations 
\begin{equation}  \label{eq:pauli-cont}
  \frac{\partial n}{\partial t} + \mathbf{\nabla}\cdot(n\mathbf{v}) = 0
\end{equation}
and 
\begin{eqnarray} 
&&\!\!\!\!\!\!\!
  m_e\frac{d\mathbf{v}}{d t} = -e (\mathbf{E}
  + \mathbf{v}\times\mathbf{B})  + \frac{\hbar^2}{2m_e}\nabla\left( \frac{\nabla^2\sqrt{n}}{\sqrt{n}} \right)
\nonumber \\ && \qquad\quad
  - \frac{2\mu_B}{\hbar}({\nabla}\otimes\mathbf{B}%
  )\cdot\mathbf{s}  - \frac{1}{%
  m_en}{\nabla}\cdot\left(n%
  \bm{\mathsf{\Sigma}} \right)
\label{eq:pauli-mom}
\end{eqnarray}
respectively. The spin contribution to Eq.\ (\ref{eq:pauli-mom}) is consistent with the results
of Ref.\ \refcite{degroot-suttorp}. Here the velocity is defined by 
\begin{equation}
\mathbf{v} = \frac{1}{m_e}\left( {\nabla}S -
i\hbar\varphi^{\dag}\mathbf{\nabla}\varphi \right) + \frac{e\mathbf{A}}{m_ec} ,
\end{equation}
the spin density vector is 
\begin{equation}
\mathbf{s} = \frac{\hbar}{2}\varphi^{\dag}\mathbf{\sigma}%
\varphi ,
\end{equation}
which is normalized according to 
\begin{equation}
  |\mathbf{s}| = \hbar/2 , 
\end{equation}  
and we have defined the
symmetric gradient spin tensor 
\begin{equation}
\bm{\mathsf{\Sigma}} = (\mathbf{\nabla}{s}_a)\otimes(%
\mathbf{\nabla}{s}^a) .
\end{equation}
Moreover, contracting Eq.\ (\ref{eq:pauli}) by $\psi^ {\dag}%
\mathbf{\sigma}$, we obtain the spin evolution equation 
\begin{eqnarray}  \label{eq:pauli-spin}
  \frac{d\mathbf{s}}{dt} = \left\{ \frac{2\mu_B}{\hbar}\mathbf{B}
    - \frac{1}{m_en}\left[\partial_a(n\partial^a\mathbf{s}) \right]\right\}\times\mathbf{s}.
\end{eqnarray}
We note that the last equation allows for the introduction
of an effective magnetic field $\mathbf{B}_{\rm eff} \equiv 
({2\mu_B}/{\hbar})\mathbf{B} - (m_en)^{-1}\left[\partial_a(n\partial^a\mathbf{s}) \right]$. However, this
will not pursued further here (for a discussion, see Ref.\ \refcite{holland}).

Comparing the effects due to spin from the Pauli dynamics with the Schr\"odinger
theory, we see a significant increase in the complexity of the fluid like equations 
due the presence of spin. The fact that the spin couples linearly to the magnetic field makes
the dynamical aspects of such Pauli systems very rich. Moreover, when going over to 
the collective regime, the back reaction through Maxwell's equation can yield interesting
new properties of such spin plasmas. In fact, the introduction of an intrinsic magnetization
can give rise to linear instability regimes, much like the Jeans instability (see Sec.\ 5.2.).

\section{Collective plasma dynamics}
As pointed out in the previous section, the route from 
single wavefunction dynamics to collective effects introduces 
a new complexity into the system. At the classical level, the ordinary pressure 
is such an effect. In the quantum case, a similar term, based on the
thermal distribution of spins, will be introduced. 

\subsection{Multistream model}
The multistream model of classical plasmas was successfully introduced
by Dawson \cite{dawson}. 
Here we will focus on the electrostatic interaction between a multistream
quantum plasma described within the Schr\"odiner model, a system first investigated in
Ref.\ \refcite{haas-etal1} (where also the stationary regime was probed). Thus, we have the governing equations
(\ref{eq:schrod-cont}) and (\ref{eq:schrod-mom}) but for $N$ beams of electrons on a stationary
ion background, \textit{i.e.},
Using this decomposition in Eq.\ (\ref{eq:schrodinger}), we obtain
\begin{equation}\label{eq:schrod-cont-multi}
  \frac{\partial n_{\alpha}}{\partial t} + \nabla\cdot(n_{\alpha}\mathbf{v}_{\alpha}) = 0 , 
\end{equation}
and 
\begin{equation}\label{eq:schrod-mom-multi}
  m_e\frac{d\mathbf{v}_{\alpha}}{d t} = e\nabla\phi
    + \frac{\hbar^2}{2m_e}\nabla\left(\frac{\nabla^2\sqrt{n_{\alpha}}}{\sqrt{n_{\alpha}}}\right) ,
\end{equation} 
now coupled through the self-consistent electrostatic potential governed by
\begin{equation}
  \nabla^2\phi = \frac{e}{\epsilon_0}\sum_{\alpha = 1}^N(n_{\alpha} - n_0).
\end{equation}
Here, $d/dt = \partial_t + \mathbf{v}_{\alpha}\cdot\nabla$ and $n_0$ is the 
density of the stationary ion background. 

In the one-stream case ($\alpha = 1$), we have the equilibrium solution $\mathbf{v} = \mathbf{v}_0$ (a constant drift relative the stationary ion background) and the constant electron density $n = n_0$ (such that $\phi = 0$).
Perturbing this system a Fourier decomposing the perturbations, such that $n = n_0 + \delta n\exp[i(\mathbf{k}\cdot\mathbf{x} - \omega t)])$, $\mathbf{v} = \mathbf{v}_0 + \delta\mathbf{v}\exp[i(\mathbf{k}\cdot\mathbf{x} - \omega t)]$, and $\phi = \delta\phi\exp[i(\mathbf{k}\cdot\mathbf{x} - \omega t)]$, we obtain
\cite{Pines,haas-etal1}
\begin{equation}
  (\omega - \mathbf{k}\cdot\mathbf{v}_0)^2 = \omega_p^2 + \frac{\hbar^2k^4}{4m_e^2} ,
\end{equation}
where the last term is the Bohm--de Broglie correction to the dispersion relation. Here we have the 
electron plasma frequency $\omega_p = (e^2n_0/\epsilon_0m_e)^{1/2}$. 

Similarly to the one-stream case, we obtain the dispersion relation \cite{haas-etal1,anderson-etal}
\begin{eqnarray}
  1 &=& \frac{\omega_{p1}^2}{(\omega - \mathbf{v}_{01}\cdot\mathbf{k})^2 -
  \hbar^2k^4/4m_e^2} \nonumber \\ 
   &&\qquad  + \frac{\omega_{p2}^2}{(\omega - \mathbf{v}_{02}\cdot\mathbf{k})^2 -
  \hbar^2k^4/4m_e^2} \ ,
\label{eq:twostreamfluid} 
\end{eqnarray}
for two propagating electron beams (with velocities $\mathbf{v}_{01}$ and $\mathbf{v}_{02}$) with background densities $n_{01}$ and $n_{02}$. The quantum
effect has a subtle influence on the stability of the perturbed plasma. For the case $n_{01} = n_{02} = n_0/2$ and $\mathbf{v}_{01} = -\mathbf{v}_{02} = \mathbf{v}_0$, we have the instability condition 
\begin{equation}
  \frac{4}{{K}^2}\left( 1 -
  \frac{1}{{K}^2} \right) < H^2 
  < \frac{4}{{K}^2}  ,
\end{equation}
in terms of the normalized wavenumber $K = kv_0/\omega_p$ and the quantum parameter 
$H = \hbar\omega_p/m_ev_0^2$ (see Fig.\ 1) \cite{haas-etal1,anderson-etal}. We see that 
when $H = 0$, we have unstable perturbations for $0 < K < 1$, but when $H \neq 0$ a considerably more
complex instability region develops. 

\begin{figure}[t]
  \centering\includegraphics[width=.7\columnwidth]{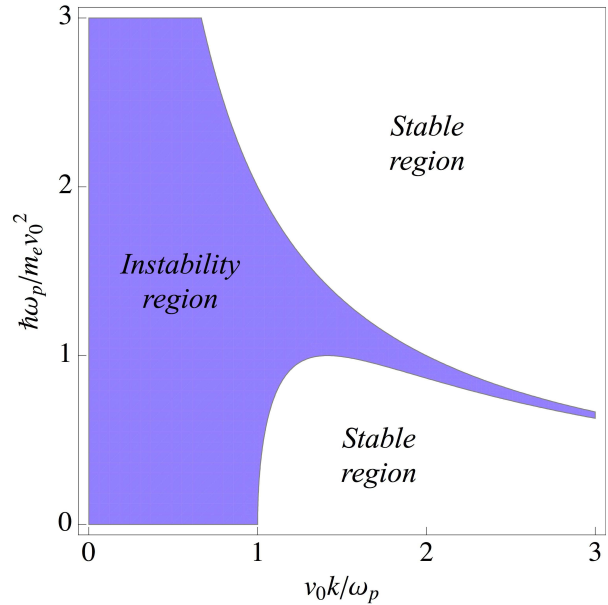}
  \caption{The regions of stability and instability in the case of the quantum two-stream
  interaction \cite{haas-etal1,anderson-etal}.}
\end{figure}

A model for treating partial coherence in such systems, based on the Wigner transform technique \cite{wigner,moyal,mendonca,schleich}, can also be developed \cite{anderson-etal} 
(see also Ref.\ \refcite{marklund}). Moreover, using the equations (\ref{eq:pauli-cont}) and (\ref{eq:pauli-mom}),
a similar framework may be set up for electron streams with spin properties.

\subsection{Fluid model}

\subsubsection{Plasmas based on the Schr\"odinger model}
Suppose that we have $N$ electron wavefunctions, and that the total system
wave function can be described by the factorization $\psi(\mathbf{x}_1, \mathbf{x}_2, \ldots \mathbf{x}_N) = \psi_{1}\psi_{2} \ldots \psi_{N}$. For each wave function $\psi_\alpha$, we have a corresponding probability
$\mathcal{P}_\alpha$. From this, we first define $\psi_\alpha = n_\alpha\exp(iS_\alpha/\hbar)$ and follow the steps leading to Eqs. (\ref{eq:schrod-cont}) and (\ref{eq:schrod-mom}). We now have $N$ such equations the wave functions $\{\psi_\alpha\}$. Defining \cite{Manfredi2005}
\begin{equation}\label{eq:meandensity}
  n \equiv \sum_{\alpha = 1}^N \mathcal{P}_\alpha n_\alpha 
\end{equation}
and 
\begin{equation}\label{eq:meanvelocity}
  \mathbf{v} \equiv \langle \mathbf{v}_\alpha \rangle = \sum_{\alpha = 1}^N \frac{\mathcal{P}_\alpha n_\alpha \mathbf{v}_\alpha}{n} ,
\end{equation}
we can define the deviation from the mean flow according to 
\begin{equation}
  \mathbf{w}_\alpha = \mathbf{v}_\alpha - \mathbf{v} .
\end{equation}
Taking the average, as defined by (\ref{eq:meanvelocity}), of Eqs. (\ref{eq:schrod-cont}) and (\ref{eq:schrod-mom}) and using the above variables, we obtain the quantum fluid equation
\begin{equation}\label{eq:cont}
  \frac{\partial n}{\partial t} + \nabla\cdot(n\mathbf{v}) = 0  
\end{equation}
and 
\begin{equation}\label{eq:mom-schrod}
  m_en\left(\frac{\partial}{\partial t} + \mathbf{v}\cdot\nabla\right)\mathbf{v} = en\nabla\phi - \nabla p + \frac{\hbar^2n}{2m_e}\nabla\left\langle \left( \frac{\nabla^2\sqrt{n_\alpha}}{\sqrt{n_\alpha}}\right) \right\rangle ,
\end{equation}
where we have assumed that the average produces an isotropic pressure $p = m_en\langle |\mathbf{w}_\alpha|^2\rangle$
We note that the above equations still contain an explicit sum over the electron wave functions. For typical scale lengths larger than the Fermi wavelength $\lambda_F $, we may approximate the last term by the Bohm--de Broglie potential \cite{Manfredi2005}
\begin{equation}
  \left\langle \frac{\nabla^2\sqrt{n_\alpha}}{\sqrt{n_\alpha}} \right\rangle \approx  
  \frac{\nabla^2\sqrt{n}}{\sqrt{n}} .
\end{equation}
Using a classical or quantum model for the pressure term, we finally have a quantum fluid 
system of equations. For a self-consistent potential $\phi$ we furthermore have
\begin{equation}  
  \nabla^2\phi = \frac{e}{\epsilon_0}(n - n_i) .
\end{equation}

\subsubsection{Spin plasmas}

The collective dynamics of electrons with spin and some of the spin 
modifications of the classical dispersion relation was presented in Ref.\ \refcite{marklund-brodin}.
Here we will follow Refs.\ \refcite{marklund-brodin} and \refcite{brodin-marklund} for the derivation
of the governing equations. 
Suppose that we have $N$ wave functions for the electrons with  
magnetic moment $\mu_e = -\mu_B$, and that, as in the case of 
the Schr\"odinger description, the total system
wave function can be described by the factorization $\psi = \psi_{1}%
\psi_{2} \ldots \psi_{N}$. Then the density is defined as in
Eq.\ (\ref{eq:meandensity}) and the average fluid velocity defined by (\ref{eq:meanvelocity}).
However, we now have one further fluid variable, the spin vector, and accordingly we let
$\mathbf{S} = \langle\mathbf{s}_\alpha\rangle $. From this we can define the microscopic
microscopic spin density $\bm{\mathcal{S}}_{\alpha} = \mathbf{s}_{\alpha} 
- \mathbf{S}$, such that $\langle\bm{\mathcal{S}}_{\alpha} \rangle
= 0$.

Taking the ensemble average of Eqs.\ (\ref{eq:pauli-cont}) we obtain the continuity equation
(\ref{eq:cont}), while we the the ensemble average applied to (\ref{eq:pauli-mom}) yield
\begin{equation}
  m_en\left( \frac{\partial }{\partial t}+\mathbf{v}\cdot {\nabla}\right) %
  \mathbf{v} = -en\left( \mathbf{E}+\mathbf{v}\times \mathbf{B}\right)  
  -\mathbf{\nabla}p 
  + \frac{\hbar^2n}{2m_e}\nabla\left( \frac{\nabla^2\sqrt{n}}{\sqrt{n}}\right) + \mathbf{F}_{\text{spin}}  
  \label{eq:mom-pauli}
\end{equation}
and the average of Eq.\ (\ref{eq:pauli-spin}) gives  
\begin{equation}
  n\left( \frac{\partial }{\partial t}+\mathbf{v}\cdot \mathbf{\nabla}\right) %
  \mathbf{S}= \frac{2\mu_B n}{\hbar }\mathbf{B}\times \mathbf{S}-\mathbf{\nabla}\cdot {%
  \bm{\mathsf{K}}} +\mathbf{\Omega}_{\text{spin}}  
\label{eq:spin}
\end{equation}
respectively. Here the force
density due to the electron spin is   
\begin{eqnarray}
  &&\mathbf{F}_{\text{spin}} = -\frac{2\mu_B n}{\hbar }(\mathbf{\nabla}\otimes \mathbf{B})\cdot %
  \mathbf{S}-\frac{1}{m_e} {\nabla}%
  \cdot \Big[ n\big(\bm{\mathsf{\Sigma}} + \widetilde{\bm{\mathsf{\Sigma}}}\,\big)\Big]  \notag \\
&&\qquad \qquad -\frac{1}{m_e}\mathbf{\nabla}\cdot \big[n(\mathbf{\nabla}%
  S_{a})\otimes \langle \mathbf{\nabla}\mathcal{S}_{\alpha }^{a}\rangle
  +n\langle \mathbf{\nabla}\mathcal{S}_{\alpha }^a \rangle \otimes (%
  \mathbf{\nabla}S_{a})\big],
\end{eqnarray}
consistent with the results in Ref.\ \refcite{degroot-suttorp}, 
while the asymmetric thermal-spin coupling is 
\begin{equation}
  \bm{\mathsf{K}} = n\langle \mathbf{w}_{\alpha }\otimes \bm{\mathcal{S}}_{\alpha}\rangle
\end{equation}
and the nonlinear spin fluid correction is 
\begin{eqnarray}
  &&\mathbf{\Omega}_{\text{spin}} = \frac{1}{m_e}\mathbf{S}\times \lbrack \partial
  _{a}(n\partial ^{a}\mathbf{S})]+\frac{1}{m_e}\mathbf{S}\times \lbrack \partial
  _{a}(n\langle \partial ^{a}\bm{\mathcal{S}}_{\alpha }\rangle )]  \notag
\\
&&\quad +\frac{n}{m_e}\left\langle \frac{\bm{\mathcal{S}}_{\alpha}}{%
  n_{\alpha}}\times \left\{ \partial _{a}[n_{\alpha}
  \partial ^{a}(\mathbf{S} + \bm{\mathcal{S}}_{\alpha })]\right\} \right\rangle 
\end{eqnarray}
where $\bm{\mathsf{\Sigma}}=(\mathbf{\nabla}S_{a})\otimes (\mathbf{\nabla}S^{a})$
is the nonlinear spin correction to the classical momentum equation, $%
\widetilde{\bm{\mathsf{\Sigma}}}=\langle (\mathbf{\nabla}\mathcal{S}_{(\alpha
)a})\otimes (\mathbf{\nabla}\mathcal{S}_{(\alpha )}^{a})\rangle $ is a pressure
like spin term (which may be decomposed into trace-free part and trace), and $[(\mathbf{\nabla}%
\otimes \mathbf{B})\cdot \mathbf{S}\,]^{a}=(\partial ^{a}B_{b})S^{b}$. 
Here the indices $a,b,\ldots = 1,2,3$ denotes the Cartesian components 
of the corresponding tensor.
We note that, apart from the additional spin density evolution equation (\ref{eq:spin}), the 
momentum conservation equation (\ref{eq:mom-pauli}) 
is considerably more complicated compared to the Schr\"odinger case represented by (\ref{eq:mom-schrod}).
Moreover, Eqs.\ (\ref{eq:mom-pauli}) and  (\ref{eq:spin}) still contains the explicit sum over the $N$
states, and has to be approximated using insights from quantum kinetic theory or some
effective theory. 

The coupling between the quantum plasma species is mediated by the
electromagnetic field. By definition, we let $\mathbf{H} = \mathbf{B}/\mu_0 - \mathbf{M}$
where $\mathbf{M} = -2n\mu_B\mathbf{S}/\hbar$ is the magnetization due to the spin sources. 
Amp\`ere's law $\mathbf{\nabla}\times\mathbf{H} = \mathbf{j} + \epsilon_0\partial_t\mathbf{E}$ 
takes the form 
\begin{equation}
  \mathbf{\nabla} \times \mathbf{B}=\mu _{0}(\mathbf{j} + \mathbf{\nabla} \times
\mathbf{M}) + \frac{1}{c^2}\frac{\partial\mathbf{E}}{\partial t},  
\label{Eq-ampere}
\end{equation}
where $\mathbf{j}$ is the free current contribution
The system is closed by Faraday's law 
\begin{equation}
  \mathbf{\nabla} \times \mathbf{E}=-\frac{\partial\mathbf{B}}{\partial t} . \label{Eq-Faraday}
\end{equation}

\section{The magnetohydrodynamic limit}

The concept of a magnetoplasma was first
introduced in the pioneering work \refcite{alfven} by Alfv\'en, who showed the existence of waves in
magnetized plasmas. Since then, magnetohydrodynamics (MHD) has found
applications in a vast range of fields, from solar
physics and astrophysical dynamos, to fusion plasmas and dusty laboratory
plasmas.

Magnetic fields, an essential component in the MHD description of plasmas,
also couples directly to the spin of the electron. Thus, the presence of spin alters the single electron
dynamics, introducing a correction to the Lorentz force term. Indeed, from the experimental perspective, a certain interest has been
directed towards the relation of spin properties to the classical theory of
motion (see, e.g., Refs.\ \refcite
{halperin-hohenberg,balatsky,rathe-etal,hu-keitel,arvieu-etal,aldana-roso,walser-keitel,qian-vignale,walser-etal,roman- etal,liboff,fuchs-etal,kirsebom-etal}%
). In particular, the effects of strong fields on single particles with spin
has attracted experimental interest in the laser community \cite
{rathe-etal,hu-keitel,arvieu- etal,aldana-roso,walser-keitel,walser-etal}.
However, the main objective of these studies was single particle dynamics, 
relevant for dilute laboratory systems, whereas our focus will be on
collective effects.
 
We will now include if the ion species, which are assumed to be described by the
classical equations and have charge $Ze$, we may derive a set of one- fluid
equations \cite{brodin-marklund}. The ion equations read 
\begin{equation}
\frac{\partial n_{i}}{\partial t}+\mathbf{\nabla}\cdot (n_{i}\mathbf{v}_{i})=0,
\label{eq:ion-density}
\end{equation}
and 
\begin{equation}
m_{i}n_{i}\left( \frac{\partial }{\partial t}+\mathbf{v}_{i}\cdot \mathbf {\nabla}%
\right) \mathbf{v}_{i}=Zen_{i}\left( \mathbf{E}+\mathbf{v}_{i}\times \mathbf{B} \right) -%
-\mathbf{\nabla}p_{i}  .  \label{eq:ion-mom}
\end{equation}
Next we define the total mass density $\rho \equiv (m_{e}n + m_{i}n_ {i})$%
, the centre-of-mass fluid flow velocity $\mathbf{V}\equiv (m_{e}n \mathbf{v}%
_{e}+m_{i}n_{i}\mathbf{v}_{i})/\rho $, and the current density $\mathbf{j}=- en%
\mathbf{v}_{e}+Zen_{i}\mathbf{v}_{i}$. Using these denfinitions, we immediately
obtain 
\begin{equation}
\frac{\partial \rho }{\partial t}+\mathbf{\nabla}\cdot (\rho \mathbf{V})=0,
\label{eq:mhd-cont}
\end{equation}
from Eqs.\ (\ref{eq:cont}) and (\ref{eq:ion-density}). Assuming
quasi-neutrality, i.e.\ $n \approx Zn_{i}$, the momentum conservation
equations (\ref{eq:mom-pauli}) and (\ref{eq:ion-mom}) give 
\begin{equation}
  \rho \left( \frac{\partial }{\partial t}+\mathbf{V}\cdot \mathbf{\nabla} \right) %
  \mathbf{V}=\mathbf{j}\times \mathbf{B}-\mathbf{\nabla}\cdot \bm{\mathsf{\Pi}}-\mathbf {\nabla}p
  + \frac{Z\hbar^2\rho}{2m_em_i}\nabla\left( \frac{\nabla^2\sqrt{\rho}}{\sqrt{\rho}}\right)  
  + \mathbf{F}_{\text{spin}},  \label{eq:mhd-mom}
\end{equation}
where $\mathbf{\mathsf{\Pi}}$ is the tracefree pressure tensor in the
centre-of-mass frame, and $P$ is the scalar pressure in the centre-of-mass
frame. We also note that due to quasi-neutrality, we have $n_
{e} \approx Z\rho /m_{i}$ and $\mathbf{v} = \mathbf{V} - m_{i}\mathbf{j}/Ze\rho $, and
we can thus express the quantum terms in terms of the total mass density $%
\rho $, the centre-of-mass fluid velocity $\mathbf{V}$, and the current $\mathbf{j}$%
. With this, the spin transport equation (\ref{eq:spin}) reads 
\begin{equation}
\rho \left( \frac{\partial }{\partial t}+\mathbf{V}\cdot \mathbf{\nabla} \right) %
\mathbf{S}=\frac{m_{e}}{Ze}\mathbf{j}\cdot \mathbf{\nabla}\mathbf{S}+\frac{2\mu_B \rho } {%
\hbar }\mathbf{B}\times \mathbf{S}- \frac{m_{i}}{Z}\mathbf{\nabla}
\cdot {\bm{\mathsf{K}}}+ \frac{m_{i}}{Z} \mathbf{\Omega}_{\text{spin}}.  \label{eq:mhd-spin}
\end{equation}

In the momentum equation (\ref{eq:mhd-mom}), neglecting the pressure and the Bohm--de Broglie potential for the sake of clarity, we have the force density $\mathbf{j}\times\mathbf{B} + \mathbf{F}_{\text{spin}}$. In general, for a magnetized medium
with magnetization density $\mathbf{M}$, Amp\`ere's law gives the free current in a finite volume $V$ 
according to 
\begin{equation}\label{eq:free-current}
  \mathbf{j} = \frac{1}{\mu_0}\mathbf{\nabla}\times\mathbf{B} - \mathbf{\nabla}\times\mathbf{M}  ,
\end{equation}
where we have neglected the displacement current. The surface current is an important part of the total current when we are interested
in the forces on a finite volume, as was demonstrated in Ref.\ \refcite{brodin-marklund} and will be 
shown below. 

It it worth noting that the expression of the force density in the momentum conservation equation 
can, to lowest order in the spin, be derived on general macroscopic grounds. 
Formally, the total force density on a 
volume element $V$ is defined as $\mathbf{F} 
= \lim_{V \rightarrow 0}(\sum_{\alpha} \mathbf{f}_{\alpha}/V)$,
where $\mathbf{f}_{\alpha}$ are the different forces acting on the volume element, and might
include surface forces as well. For 
magnetized matter, the total force on an element of volume $V$ is then
\begin{equation}
  \mathbf{f}_{\rm tot} = \int_V\mathbf{j}_{\rm tot}\times\mathbf{B}\,\mathrm{d}V 
    + \oint_{\partial V}(\mathbf{M}\times\hat{\mathbf{n}})\times\mathbf{B}\,\mathrm{d}S 
\end{equation}
where (neglecting the displacement current) $\mathbf{j}_{\rm tot} = \mathbf{j} + \mathbf{\nabla}\times\mathbf{M}$.
Inserting the expression for the total current into the volume integral and using the divergence
theorem on the surface integral, we obtain the force density
\begin{equation}
  \mathbf{F}_{\rm tot} = \mathbf{j}\times\mathbf{B}
    + M_k\mathbf{\nabla}B^k  ,
\end{equation}
identical to the lowest order description from the Pauli equation (see Eq.\ (\ref{eq:mhd-mom})). 
Inserting the free current expression (\ref{eq:free-current}), due to Amp\`ere's law, we can write
the total force density according to
\begin{equation}\label{eq:total-force}
   {F}^i = -\partial^i\left( 
    \frac{B^2}{2\mu_0} - \mathbf{M}\cdot\mathbf{B} 
  \right) + \partial_k({H^iB^k}) .
\end{equation}
The first gradient term in Eq.\ (\ref{eq:total-force}) can be interpreted as the 
force due to a potential (the energy of the magnetic field and the magnetization
vector in that field), while the second divergence term is the anisotropic magnetic pressure effect. 
Noting that the spatial part of the stress tensor takes the form \cite{degroot-suttorp} 
\begin{equation}
  T^{ik} = -H^iB^k +(B^2/2\mu_0 - \mathbf{M}\cdot\mathbf{B} )\delta^{ik} , 
\end{equation}
we see that the
total force density on the magnetized fluid element can be written $F^i = -\partial_kT^{ik}$, 
as expected. Thus, the Pauli theory results in the same type of conservation laws as
the macroscopic theory. 
The momentum conservation equation (\ref{eq:mhd-mom}) then reads
\begin{equation}
  \rho \left( \frac{\partial }{\partial t}+\mathbf{V}\cdot \mathbf{\nabla} \right)\mathbf{V} = %
  -\mathbf{\nabla}\left( 
    \frac{B^2}{2\mu_0} - \mathbf{M}\cdot\mathbf{B} 
  \right) + B^k\partial_k\mathbf{H}
  -\mathbf {\nabla}p
  ,  \label{eq:mhd-mom2}
\end{equation}
where for the sake of clarity we have assumed an isotropic pressure, dropped the displacement
current term in accordance with the nonrelativistic assumption, and neglected the
Bohm potential (these terms can of course simply be added to (\ref{eq:mhd-mom2})). 
This concludes the discussion of the spin-MHD plasma case. Next, we will look at some applications
of the derived equations. However, it should be noted that in many cases the spins are close
to thermodynamic equilibrium, and we can thus write the paramagnetic electron response
in terms of the magnetization \cite{brodin-marklund}
\begin{equation}
  \mathbf{M}=\frac{\mu _{B}\rho}{m_i}\,\tanh \left( \frac{\mu _{B}B}{k_BT}\right)
  \widehat{\mathbf{B}} , \label{Eq: tanh-factor}
\end{equation}
instead of using the full spin dynamics. 
Here $B$ denotes the magnitude of the magnetic field and $\widehat{\mathbf{B}}$
is a unit vector in the direction of the magnetic field, $k_B$ is Boltzmann's constant, and $T$ is the electron temperature.

\section{Examples and applications}
The above equations are quite complicated, but as such also extremely
rich. Suitable and physically relevant approximations, such as the magnetization given by the expression (\ref{Eq: tanh-factor}), will however lead to considerable simplifications. Below we consider
two specific examples where such simplifying assumptions lead to interesting spin effects.

\subsection{Spin solitons}
In Ref.\ \refcite{BM-pairplasma}, it was shown that the electron spin
can introduce novel nonlinear structures in plasmas, with no limiting classical 
counterpart. In particular, the MHD limit for a \emph{electron-positron pair plasma} is considered. 

Neglecting dissipative effect, the governing equations for the system of interest read
\begin{equation}
\frac{\partial \rho }{\partial t}+\mathbf {\nabla}\cdot (\rho \mathbf{v})=0,
\end{equation}
\begin{equation}
\rho \left( \frac{\partial }{\partial t}+\mathbf{V}\cdot \mathbf{\nabla}\right) %
\mathbf{V}=\mathbf{j}\times \mathbf{B}-\mathbf{\nabla}\cdot \bm{\mathsf{\Pi}}
+ \frac{ \hbar ^{2}\rho}{2m_e^{2}} \mathbf{\nabla}\left( \frac{\nabla ^{2}\sqrt{\rho}}{\sqrt{\rho}}\right) 
+\mathbf{F}_{\text{spin}},
\end{equation}
where $\bm{\mathsf{\Pi}}=[(T_{e}+T_{p})/2m_{e}]\bm{\mathsf{I}}+(m^{2}/\rho )%
\mathbf{j}\otimes \mathbf{j}$ is the total pressure tensor in the centre-of-mass
frame and 
\begin{eqnarray}
  \mathbf{F}_{\text{spin}}=
 2\tanh
  \left( \frac{\mu _{B}B}{k_BT}\right) \frac{\mu _{B}}{m_e}\mathbf{\nabla}B ,
\label{eq:quantum2}
\end{eqnarray}
where we have assumed equal temperature $T$ of the electrons and positrons. 
Moreover, we have 
\begin{equation}
\frac{\partial \mathbf{B}}{\partial t}=\mathbf{\nabla}\times \left( \mathbf{V}\times %
\mathbf{ B}\right) ,  \label{eq:simple-ohm}
\end{equation}
while the current is given by 
\begin{equation}
\mathbf{j} = \frac{1}{\mu _{0}}\nabla \times \mathbf{B} - \epsilon _{0}\frac{%
\partial }{\partial t}\left( \mathbf{V}\times \mathbf{ B}\right) - \frac{\mu _{B}}{2m}\mathbf{\nabla}\times \left[ \rho \tanh \left( 
\frac{\mu _{B}B}{k_BT}\right) \mathbf{\hat{B}}\right]
\label{eq:Ampere-Ohm}
\end{equation}
For one-dimensional Alfv\'en waves, the above system can be reduces to the modified
Korteweg-de Vries equation\cite{BM-pairplasma}
\begin{eqnarray}
  \left[ \frac{\partial }{\partial t}+c_{A,\mathrm{sp}}\cos \theta \frac{%
  \partial }{\partial \xi }+\frac{c_{A}^{3}\cos ^{3}\theta }{2\omega
  _{c}^{2}\sin ^{2}\theta }\frac{\partial ^{3}}{\partial \xi ^{3}}\right]
  V_{y}
  = - \frac{(\mu _{B}B_{0})^2}{2m_ek_BT}\frac{V_{y}^{2}}{%
  c_{A}^{3}}\frac{\partial V_{y}}{\partial \xi }  \label{eq:ALf-NL2}
\end{eqnarray}
where $c_{A}=\left[ c^{2}B_{0}^{2}/(c^{2}\mu _{0}\rho _{0}+B_{0}^{2})\right] ^{1/2}$ 
is the Alfv\'en speed, $c_{A,\mathrm{sp}}=c_{A}^{2}/(1+\delta _{\rm sp})$ is the spin-modified
Alfv\'{e}n speed, $\omega_c = eB_0/m_e$ is the cyclotron frequency, 
$B_{0}$ is the magnitude of the unperturbed magnetic field, $\rho
_{0}$ is the unperturbed density, and we have the spatial coordinate $\xi =x\sin \theta +z\cos \theta $. 
We see that neglecting the spin contribution leads to a purely dispersive equations. Thus,
the spin enables the formation of solitons with no limiting classical solution.

\subsection{Ferromagnetic plasma behaviour}

For an ion-electron plasma, we have the governing equations\cite{brodin-marklund2}
\begin{equation}
\frac{\partial \rho }{\partial t}+\bm {\nabla}\cdot (\rho \bm{V})=0,
\end{equation}
the momentum equation
\begin{eqnarray}
&&\!\!\!\!\!\!\!\!\!\!\!\! \rho \left( \frac{\partial }{\partial t}+\bm{V}\cdot \bm{\nabla}\right) %
\bm{V}=-\bm{\nabla}\left( \frac{B^{2}}{2\mu _{0}}-\bm{M}\cdot \bm{B}\right)  -\bm {\nabla}p
\nonumber \\ &&\!\!\!\!
+ \bm{B}\cdot \bm {\nabla}\left( \frac{1}{\mu_0}\bm{B}-\bm{M}\right) + \frac{%
\hbar^2\rho}{2m_em_i}\bm{\nabla}\left( \frac{\nabla ^{2}\sqrt{\rho}}{\sqrt{%
\rho}}\right)\, ,  
\end{eqnarray}
and the idealized Ohm's law
\begin{equation}
\frac{\partial \bm{B}}{\partial t}=\bm{\nabla}\times \left( \bm{V}\times %
\bm{B}\right)  , \label{Eq:Ohm}
\end{equation} 
where the variables are defined as above. Using the magnetization (\ref{Eq: tanh-factor}),
we obtain a closed set of equations.

In what follows, we will study the linear modes of this system, with a
particular focus on the stability properties. With $\rho = \rho_{0} + \rho_{1}$, $\bm{B}=%
\bm{B}_{0}+\bm{B}_{1}$, $\bm{M}=\bm{M}_{0}+\bm{M}_{1}$, and $\bm{v} = \bm{v}%
_1$, such that $\rho_1 \ll \rho_0$, $|\bm{B}_1| \ll |\bm{B}_0|$, $|\bm{M}_1|
\ll |\bm{M}_0|$, and $\bm{B}_0 = B_0\hat{\bm{z}}$, we linearize our equations in the perturbed variables.
Assuming that the background quantities are constants, the general
dispersion relation can, after a Fourier decomposition, be written
\begin{equation}
\left( \omega ^{2}-k_{z}^{2}\widetilde{C}_{A}^{2}\right) \bigg[ \left(
\omega ^{2}-k^{2}\widetilde{C}_{A}^{2}-k_{x}^{2}\widetilde{V}_{A}^{2}(k)\right) \left(
\omega ^{2}-k_{z}^{2}V_{A}^{2}(k)\right)
+k_{x}^{2}k_{z}^{2}\widetilde{V}_{A}^{4}(k)%
\bigg] =0  \label{Eq:General-DR}
\end{equation}
where $\widetilde{C}_{A}$ is the spin-modified Alfv\'{e}n velocity given by
\begin{equation}
\widetilde{C}_{A}=\frac{C_{A}}{\left[ 1+ (\hbar \omega
_{pe}^{2}/2m_ec^{2}\omega _{ce}^{(0)})\tanh (\mu _{B}B_{0}/k_BT)\right] ^{1/2}}
\,\, ,  \label{Eq-Alf-spin-velocity}
\end{equation}
$C_{A}$ is the standard Alfv\'{e}n velocity $C_{A}=(B_{0}^{2}/\mu _{0}\rho
_{0})^{1/2}$,
\begin{equation}
\widetilde{V}_{A}^{2}(k)=V_{A}^{2}(k)-\frac{\hbar \omega _{ce}}{m_{i}}\tanh \left(
\frac{\mu _{B}B_{0}}{k_BT}\right) \, ,
\end{equation}
and
\begin{equation}
V_{A}^{2}(k)=c_{s}^{2}+\frac{\hbar ^{2}k^{2}}{4m_{i}m_{e}}
\end{equation}
Here $\omega _{pe}=(\rho_{0}e^{2}/\varepsilon _{0}m_em_i)^{1/2}$ is the plasma
frequency, $\omega _{ce}^{(0)}$ is the electron cyclotron frequency associated with
the external magnetic field (i.e. with the contribution to $B_{0}$ from the
spin sources excluded). The relation between the full electron
cyclotron frequency $\omega _{ce}=eB_{0}/m_e$ and
$\omega _{ce}^{(0)}$ is given by $\omega_{ce}=\omega _{ce}^{(0)} + (\hbar
\omega _{pe}^{2}/m_ec^{2})\tanh (\mu _{B}B_{0}/k_BT)$. We stress that $\widetilde{V}_{A}$,
which to some extent can be considered as an effective acoustic velocity,
may be imaginary for a strongly magnetized plasma due to the spin
contribution, a fact which will be explored in some detail below.

We
consider propagation perpendicular to the external magnetic field, which is
the geometry which leads to instability most easily. For the case $\bm{k}=k%
\hat{\bm{x}}$, Eq. (\ref{Eq:General-DR}) reduces to
\begin{eqnarray}
&&\!\!\!\!\!\!\!\!\!\!\!\! \omega =k\Bigg[ \frac{C_{A}^{2}}{1+(\hbar \omega _{pe}^{2}/2m_ec^{2}\omega
_{ce}^{(0)})\tanh (\mu _{B}B_{0}/k_BT)}
\nonumber \\ && \quad\qquad
+c_{s}^{2}+\frac{\hbar ^{2}k^{2}}{%
4m_{i}m_{e}}-\frac{\hbar \omega _{ce}}{m_{i}}\tanh \left( \frac{\mu _{B}B_{0}%
}{k_BT}\right) \Bigg] ^{1/2}  \label{Eq:Instability-DR} .
\end{eqnarray}

The necessary and sufficient instability condition can thus be written as
\begin{equation}
P_{\mathrm{sp}}+P_{\mathrm{m}}+P+\frac{\rho_{0}\hbar ^{2}k^{2}}{4m_{e}m_i}<0 ,
\label{Eq:Full-condition}
\end{equation}
where the total pressure $P_{\mathrm{tot}}=P_{\mathrm{sp}}+P_{\mathrm{m}}+P$
consists of the effective spin pressure $P_{\mathrm{sp}}=-(\rho_{0}\hbar \omega
_{ce}/m_i)\tanh (\mu _{B}B_{0}/k_BT)$, which is the only negative pressure term
and therefore the source of the instability, the magnetic pressure $P_{%
\mathrm{m}} = {\rho_0C_{A}^{2}}/[{1 + (\hbar
\omega _{pe}^{2}/2m_ec^{2}\omega _{ce}^{(0)})\tanh (\mu _{B}B_{0}/k_BT)}]$, 
and the particle pressure $P=n_{0}m_{i}c_{s}^{2}$, containing
both the thermal and Fermi pressure part. Thus, a plasma can contain a magnetization instability, 
much like the gravitationally induced Jeans instability \cite{brodin-marklund2}.

\section{Conclusions and future possibilities}

We have seen that quantum effects, and in particular the electron spin, can introduce new and interesting
aspects in plasma theory and experiments. It is expected that the rapid development of quantum plasma theory will be fueled by recent experiments (see, \textit{e.g}, Ref.\ \refcite{glenzer-etal}), and that new regimes of interest will enter the
arena as the next-generation laser systems gets online. 

Some particular developments that would be of interest is to further look at spin effects from a kinetic perspective \cite{cowley-etal,kulsrud-etal}. Such treatments would be similar to the density matrix approach, and could, by using analogies from classical kinetic theory, spur the experimental interest in collective quantum effects and the transition from quantum to classical behaviour. Moreover, the fluid equations presented here has not been analyzed to their full extent. For example, there are terms which have been neglected, that could produce interesting nonlinear effects in quantum plasmas, such as spin self-interaction. Such topics will be approached in future research.


\end{document}